%% file: Main.tex
\documentclass[conference]{IEEEtran}
\IEEEoverridecommandlockouts
\usepackage{cite}
\usepackage{amsmath,amssymb,amsfonts}
\usepackage{algorithmic}
\usepackage{graphicx}
\usepackage{textcomp}

\usepackage{xcolor}
\usepackage{balance}
\usepackage[utf8x]{inputenc}

\graphicspath{{img/}{photos/}}
\def\BibTeX{{\rm B\kern-.05em{\sc i\kern-.025em b}\kern-.08em
    T\kern-.1667em\lower.7ex\hbox{E}\kern-.125emX}} 
\begin{document}

\title{Evaluating Code Metrics in GitHub Repositories Related to Fake News and Misinformation}

\author{
    \IEEEauthorblockN{Jason Duran}
    \IEEEauthorblockA{\textit{Computer Science Department} \\
    \textit{Boise State University}\\
    Boise, Idaho, USA \\
    jasonduran@u.boisestate.edu}
    \and

    \IEEEauthorblockN{ Mostofa Najmus Sakib}
    \IEEEauthorblockA{\textit{Computer Science Department} \\
    \textit{Boise State University}\\
    Boise, Idaho, USA \\
    mostofanajmussak@u.boisestate.edu}
    \and

    \IEEEauthorblockN{ Nasir U. Eisty}
    \IEEEauthorblockA{\textit{Computer Science Department} \\
    \textit{Boise State University}\\
    Boise, Idaho, USA \\
    nasireisty@boisestate.edu}
    \and
    
    \IEEEauthorblockN{Francesca Spezzano}
    \IEEEauthorblockA{\centerline{\textit{Computer Science Department}} \\
    \textit{Boise State University}\\
    Boise, Idaho, USA \\
    francescaspezzano@boisestate.edu}
   }
\maketitle

\begin{abstract}
\input{abstract.tex}
\end{abstract}

\begin{IEEEkeywords}
GitHub Repository Mining, Fake News, Misinformation, Code Metrics, Source Code Analysis
\end{IEEEkeywords}

\section{Introduction}
\label{sec:Intro}
\input{introduction.tex}

\nocite{anu2020,Daniel2020-lq,zhou2020survey}

% \section{RESEARCH GOAL AND QUESTIONS}
% \label{sec:Research_goal_and_questions}
% \input{Research_goal_and_questions.tex}

\section{Background}
\label{sec:related}
\input{related}

\section{Methodology}
\label{sec:Methodology}
\input{methodology}

%\section{Experiment}
%\label{sec:Experiment}
%\input{experiment}

\section{Results}
\label{sec:Result}
\input{result}
\begin{figure}[ht]
\centering
\centerline{\includegraphics[width=.5\textwidth]{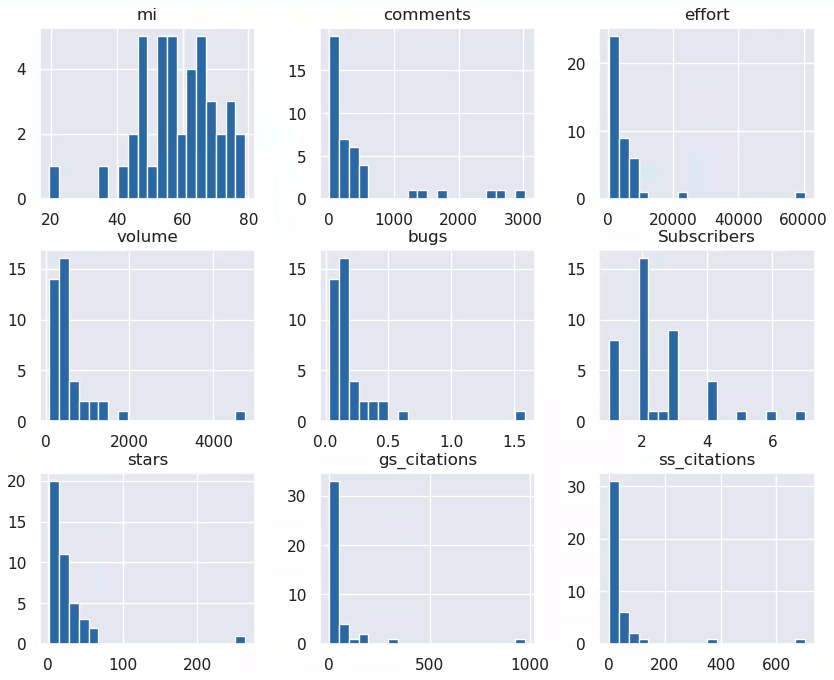}}
\caption{Histograms of Model Variables}
\label{regressors}
\end{figure}

\begin{figure}[ht]
\centering
\centerline{\includegraphics[width=.5\textwidth]{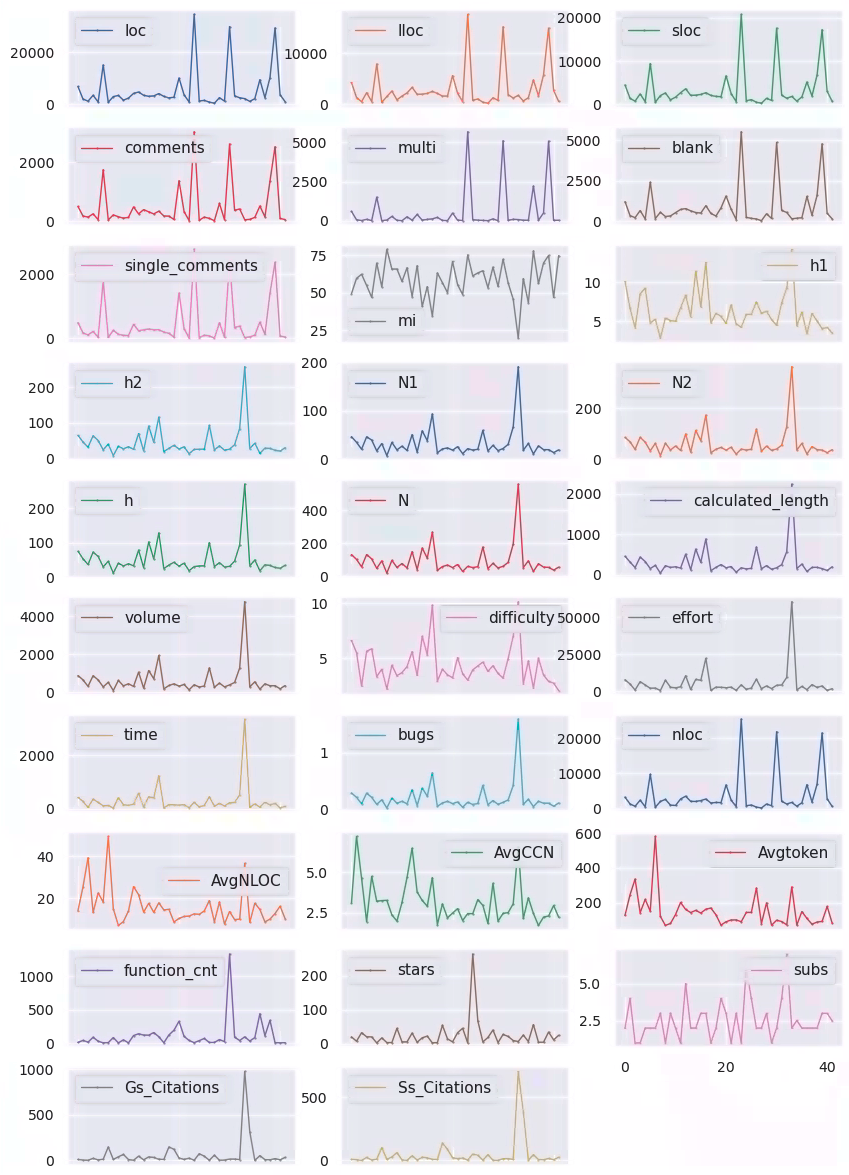}}
\caption{Graph of Values of Repository Metrics}
\label{allvariables}
\end{figure}

% \begin{figure}[ht]
% \centering
% \centerline{\includegraphics[width=.5\textwidth]{ind_vars2.png}}
% \caption{Histograms of Independent Variables}
% \label{citationshist}
% \end{figure}

\section{Threats to Validity}
\label{sec:threats}
\input{threats}

%\section{Future Work}
%\label{sec:future}
%\input{future}

\section{Discussion and Conclusion}
\label{sec:Conclusion}
\input{conclusion}

\balance

\bibliographystyle{plain}
\bibliography{Main}

\end{document}

%% file: abstract.tex
%Abstract
The surge of research on fake news and misinformation in the aftermath of the 2016 election has led to a significant increase in publicly available source code repositories. Our study aims to systematically analyze and evaluate the most relevant repositories and their Python source code in this area to improve awareness, quality, and understanding of these resources within the research community. Additionally, our work aims to measure the quality and complexity metrics of these repositories and identify their fundamental features to aid researchers in advancing the field's knowledge in understanding and preventing the spread of misinformation on social media. As a result, we found that more popular fake news repositories and associated papers with higher citation counts tend to have more maintainable code measures, more complex code paths, a larger number of lines of code, a higher Halstead effort, and fewer comments. Utilizing these findings to devise efficient research and coding techniques to combat fake news, we can strive towards building a more knowledgeable and well-informed society.

%% file: introduction.tex
%Introdeuction

In the current era, software and technology have become ubiquitous and play a pivotal role in the progress and advancement of various domains. Therefore, it's crucial to delve into software quality, applicability, and work processes. To assess the quality of software or toolkits, a direct approach is to access the source code from popular repositories such as Bitbucket, GitHub, or other version control systems and analyze them. This analysis can reveal existing software and tools' interesting, attractive, user-friendly, and reliable features.

Typically, the quality of the software is measured using code quality metrics. Simply put, metrics development involves creating standard numerical values that serve as a benchmark for evaluating quality. As the software industry has grown and become more automated, code quality metrics have become an essential tool for assessing the effectiveness of source code. This approach can provide valuable insights into the quality of work and how it relates to the characteristics of real-world samples. 

The study of popular software repositories and their code bases is an established area of research in software engineering. Similarly, a significant amount of research has focused on detecting fake news in virtual media, spanning computer science, and social science. Fake news is intentionally fabricated content that aims to deceive readers, leading to misinformation~\cite{allcott2017social}. The field has received considerable funding from the US and international entities, making it a significant focus of academic research for the foreseeable future. In addition, it is a highly cited research area that has garnered interest from researchers worldwide.  

The prevalence of fake news across social, online, and print media has raised doubts about the reliability of news and information among the general public~\cite{karimi-etal-2018-multi}. This issue has garnered significant attention from both the public and academic spheres, especially in this era of abundant information~\cite{ruchansky2017csi}. In today's society, false, misleading, and politically charged fake news stories are gaining more traction and creating distrust and skepticism among audiences, especially in the age of social media dominance. The spread of fake news and its harmful impact on society, particularly after the 2016 US presidential election, is a significant phenomenon~\cite{pennycook2018prior}.

% \begin{figure}[htp]
%     \centering
%     \includegraphics[width=8.5cm]{bi.png}
%     \caption{Dissemination of Fake news and posts in Facebook on the eve of the 2016 US election}
%     \label{fig:Fake_news}
% \end{figure}

% Figure \ref{fig:Fake_news} which is collected from online news source show a real example of the extent of fake news during the 2016 election on the social media platform Facebook. During that time, the top 20 fake news achieved even higher popularity in terms of network traffic compared to the 20 real news \cite{intro_bi_1}. Similarly, all the news outlet in the current time heavily relies on social media as a source of traffic. This opens the door for sharing and disseminating fake news on social media and mainstream news platforms.  

% Fake news has distinguished characteristics. In the paper \cite{kumar2018false}, the authors Srijan Kumar and Neil Shah presented a comprehensive study of the actors, rationale, and impact of the successful spreading of false information, as well as its characteristics of it and its detection algorithm. Shrestha and Spezzano \cite{shrestha2020multi} found that News headlines are more informative than news body content, suggesting that we can avoid “reading” the news excerpt and focus on other modalities to detect misleading news better. Horne and Adali pointed out that \cite{horne2017just}, fake news articles are shorter in terms of content but use repetitive language, less punctuation, and fewer quotes. 

In response to the growing prevalence of fake news, a significant amount of research has focused on detecting fake news and misinformation using artificial intelligence (AI) and machine learning (ML) and 
researchers worldwide have developed many software, tools, and models. For instance, Monti et al. \cite{monti2019fake} created a generic fake news detection model based on geometric deep learning. Shu et al. \cite{10.1145/3289600.3290994} established a tri-relationship between publisher-news relations and user-news interactions to classify fake and real news. Wang et al. \cite{wang2022fmfn} devised a fine-grained multimodal fusion network-based model that leverages textual and visual features to detect fake news. 

However, little effort has been dedicated to evaluating the quality and functionality of the source code of these tools using popular code metrics. To address this research gap, we analyze well-cited and famous repositories related to fake news and misinformation. By analyzing citation data, we can gain insights into the importance of ideas, changes in source work, and the impact of research in a particular field \cite{anderson2023citation}. Additionally, meta information such as stars and forks can provide insight into the effects of GitHub repositories associated with fake news. We specifically focus on exploring publicly available repositories on GitHub, as it is the largest freely available source code hosting repository for software developers~\cite{borges2016predicting}. Our study aims to determine the relationship between measures of popularity, such as citation and star count, and traditional code metrics. 
 
To summarize, our research intends to bridge the gap between software engineering and fake news, two distinct fields of computer science research, and to advance the field while also presenting fresh research opportunities. Our overarching research objective is to identify code quality and complexity metrics associated with popular fake news research and repositories. This will provide researchers with valuable insights into the aspects of their work and code that are highly valued by others. Additionally, since these repositories are highly regarded and frequently cited in the research community, it motivates us to continue exploring this topic. Therefore, we pose the following research questions:  

% \subsection{Research Goal} 

% Our research goal is to identify measures of code quality and complexity that may be associated with more popular fake news research or more popular fake news code repositories. This will ideally give researchers insight into the aspects of their research and shared code that are more valued by others.
 
%\subsection{Research Questions} 

\textbf{RQ1: Is there any correlation between different code metrics and research citations?}

The primary objective of RQ1 is to investigate whether any discernible code metrics are linked to the citations received by a paper for a particular source code repository. We will identify commonly used measures of code quality, complexity, and characteristics and examine their correlation with the citations of related research papers. It should be noted that this analysis does not necessarily indicate causality. Still, it may provide valuable insight into the aspects of shared research and source code that are highly valued.
 
\textbf{RQ2: Is there any correlation between different code metrics and GitHub popularity?}

Similar to RQ1, this research question aims to explore the association between various code metrics and measures of popularity, but this time with respect to repository-specific measures such as GitHub stars and subscribers. It is worth noting that while these measures are related to popularity, per Fig.~\ref{fig:corr} they are only loosely correlated with citations. Thus, we expect different metrics associated with repository popularity compared to paper citations. We aim to identify these metrics and gain insight into the factors that contribute to the popularity of repositories in the domain of fake news detection.

\textbf{RQ3: Is there any correlation between GitHub popularity and associated paper citations regarding code metrics?}

The goal of this research question is to investigate if there are any common code metrics or themes that are linked with the popularity of a repository based on its GitHub stars, subscribers, and paper citations. 

% \section{Research Goal and Questions}
% \subsection{Research Goal} 

% Our research goal is to provide insight into the various software repositories for fake news research available such that researchers might have more insight into what provides a better or more popular public repository. 

% \subsection{Research Questions} 

% RQ1:What are the typical code metrics of a fake news repository?\\

% Questions we look to answer in this section are what does the typical (median in our case due to outliers) fake news research repositories look like? What are average lines of code, comments, code complexity, etc.? What \% of the repositories have README.md files and requirements.txt files? \\

% RQ2: Which repositories, in particular, stand out in terms of metrics?\\

% Which repositories are the largest? Have the most complex code? The simplest code? 
% The most comments? Lines of code? What are prevalent deep learning libraries in use across 
% projects? \\

% RQ3: How do code metrics vary for more/less popular repositories?\\

% What we hope to answer with this section is how to all the above-mentioned measures relate to the popularity of the repository measured by either stars or followers. In an ideal world, we would also have downloads, views, and citations related to the research paper for the repository to use, hopefully for future work. 

%% file: related.tex
GitHub has become a popular platform for software quality analysis through source code analysis among software engineering researchers. It offers open-source capabilities, integrated social features, and metadata that can be retrieved through a simple API call~\cite{perils}. Research on GitHub can be categorized into two sections: quantitative and qualitative. Quantitative studies focus on the development practices followed by developers, while qualitative analysis evaluates the performance and quality of projects by examining developer activities~\cite{perils}.

Kalliamvakou et al.~\cite{perils} address the challenges and opportunities of using GitHub as a data source for software quality analysis in a comprehensive study. They analyze a dataset of GitHub repositories from 2014 and identify various limitations of using such data, including inactive projects, repositories used solely for disk storage, and lack of pull request usage. Finally, the authors discuss the implications of these limitations and provide recommendations for researchers to ensure the validity and reliability of their findings. 

Inspired by social media, GitHub introduced stars to measure a repository's popularity. Borges et al.~\cite{borges2016predicting} used star count as a measure of popularity. Meanwhile, Hu et al.~\cite{hu2016influence} utilized stars and forks to develop a star graph revealing the social relationship between users and repositories. Additionally, Zhu et al.~\cite{zhu2014patterns} found that the use of standard folders, such as documents, testing, and examples, increases the probability of a code repository being forked and thus directly related to popularity. Interactions among the developer community on GitHub, such as following other users, watching or starring other project repositories, can also impact exposure and popularity~\cite{dabbish2012social}. Aggarwal et al.~\cite{aggarwal2014co} explored the correlation between documentation and popularity, concluding that popularity can positively impact the attraction of more developers and the continuous improvement of documentation. Additionally, citation counts are a way to measure the popularity of a paper or repository, reflecting key concepts, approval from scholars, and groundbreaking novel ideas in a scientific domain~\cite{small1973co}.       

According to Sharma~\cite{sharma2020we}, software developers have utilized code quality metrics for over five decades. These metrics are typically divided into five categories: size, complexity, coupling, cohesion, and inheritance. Over time, code quality metrics have naturally progressed and can now be assessed using various factors, such as lines of code, code clone, readability, and author rank~\cite{ganjare2015measuring}. Among the most prominent code quality metrics are Cyclomatic complexity~\cite{mccabe}, Halstead metrics~\cite{halstead1977elements}, Chidamber and Kemerer (C\&K) metrics suite~\cite{chidamber1994metrics}, and the MOOD metrics~\cite{abreu1994object}. 

McCabe~\cite{mccabe} developed a method to measure code complexity using graph theory to analyze code paths. This measurement estimates the number of independent code paths a system can take and is considered to closely reflect the perceived complexity of a code by programmers. In brief, the calculation for McCabe code complexity is $MCC = C + 1$, where $MCC$ represents the McCabe code complexity, and $C$ denotes the number of decision points in the code~\cite{ChatGPT}.

%\begin{equation}
%MCC = C + 1
%\end{equation}

Halstead metrics consider the operators (such as main, print, int, etc.) and operands (\%d, avg, etc.) in code to determine the time required for programming, the number of delivered bugs, the level of difficulty, and the amount of effort needed, among other factors. On the other hand, the Chidamber and Kemerer (C\&K)~\cite{chidamber1994metrics} metrics suite comprises six object-oriented metric suites, namely Weighted Methods per Class, Depth of Inheritance Tree, Number of Children, Coupling between Objects, Response for a Class, and Lack of Cohesion of Methods. Finally, the MOOD metrics are a collection of metrics that include method inheritance factor, attribute inheritance factor, coupling factor, method inheritance factor, attribute inheritance factor, and coupling factor~\cite{abreu1994object}. 

During our research, we aimed to identify any metrics explicitly designed for assessing fake news-related repositories. However, to our knowledge, no such metrics have been developed. Therefore, we have resorted to utilizing some of the code quality metrics mentioned earlier to evaluate our selected repositories.

%% file: methodology.tex
In this section, we present our approach to gathering the data and the analysis methods used to examine the relationships between the code quality metrics, GitHub, and citation-related measures of popularity.

\subsection{Data Collection}

Our objective was to locate repositories specifically associated with published, submitted, or presented papers on fake news detection models. To identify these repositories, we initiated a search on GitHub for all repositories containing the keywords ``fake news" or ``misinformation" created after 1st January 2019, with a minimum of 2 subscribers and larger than 1 MB in size. We arrived at these search criteria after experimenting with various options to balance, including as many relevant repositories as possible while avoiding small test repositories, class assignments, and other less significant repositories that would consume our limited API requests. Following this process, we identified 445 repositories related to ``fake news" and 71 related to ``misinformation" as potential candidates.

We refined our analysis by requesting specific attributes from GitHub and eliminating repositories that did not relate to model-related research projects. Firstly, we excluded repositories with an associated URL containing `.ai', `.io', `.guru', `.br', `.Br', or `flask'. These URLs typically represented demo applications or advanced class or resume-building projects. Next, we included any repository that had an associated URL linking to a paper or research site and mentioned ``paper", ``conference", or ``proceedings" in either the GitHub description or the `README.md' file. Finally, we required each repository to contain at least one `.py' file for our analysis. Furthermore, we manually excluded a few repositories that caused third-party libraries to crash while parsing the Python code, mainly those written in Python 2.7 or earlier versions. From this exploration, We ended up with 39 GitHub repositories.

%We initially anticipated building a manual list of repositories to analyze but found some simple inclusion/exclusion criteria to be sufficient and robust enough to allow us to pick up new repositories and vary our date range, as our API limits allowed us to support many more samples than manually maintaining a list would allow. 
 
Apart from the aforementioned details, we searched for citations of each repository that had an associated paper. To accomplish this, we utilized two platforms, namely Google Scholar \cite{google_scholar} and Semantic Scholar \cite{semantic_scholar}. We manually examined each repository that appeared studies linked above, determined the paper title, and manually searched the websites to gather the \# of citations.  This data was then used alongside the stars and subscribers as independent variables for the regression modeling. 

We manually retrieved the source code repositories for three major works~\cite{yang2022reinforcement,10.1145/3357384.3357862, 10.1145/3292500.3330935} in the fake news domain that was not available on GitHub and added them to our analysis directory. These repositories were maintained on publicly accessible cloud solutions. To compensate for missing stars and subscribers (followers) counts, we assigned them the average values for the overall sample. We gathered citations for these repositories in the same manner as for the other repositories.

%total
After applying the inclusion and exclusion criteria mentioned earlier, we obtained a total of 42 repositories and projects. We used this dataset to conduct the planned analysis to explore the relationship between code metrics, repositories, and citations.
% \subsection{Data Analysis Methodology}

% For analyzing the data, we used several python libraries and followed the standard data cleaning process before checking software metrics from them. 

\subsection{Data Analysis}
We utilized two popular Python libraries (Radon \& Lizard) to compute code metrics on Python code, calculating several code quality, complexity, and maintainability metrics typically used in software engineering. Tables \ref{Radon} and \ref{Lizard} display these metrics and their corresponding definitions. We used the Radon~\cite{radon} and Lizard~\cite{lizard} Python packages to calculate these metrics on a per-file and function level. We then aggregated these metrics by taking the mean of the file level measures to the repository level for comparing projects and analyzing the relationship between the metrics and popularity as represented by GitHub stars and subscribers. Additionally, we calculated several associated meta-variables such as the presence of a requirement.txt file, a README.md that exceeds a certain length, and which deep learning libraries (PyTorch, Keras, and/or TensorFlow) the Python code references or uses.

Table \ref{data_statistics} shows the summary data statistics of the variables from the repositories in GitHub and the remaining projects for the calculated metrics. The last row or the maximum value of the measured metrics is particularly interesting because it shows some of the largest outliers. That is also supported in the histogram of the model variables in Fig.~\ref{regressors} and the line graphs of all the variables in Fig.~\ref{allvariables}.   

%\begin{figure}[htp]
%    \centering
%    \includegraphics[width=8.5cm]{miFormula.png}
%    \caption{Formula for MI (Maintainability Index) calculation from radon documentation~\cite{radon}}
%%\end{figure}

\begin{table}[ht]
\renewcommand{\arraystretch}{1.3}
\caption{Metrics calculated from Radon}
\label{Radon}
\begin{tabular}{|l|l|}
\hline
\textbf{Metric}          & \textbf{Description}                                                                                                                                                                                                                                                                                                                                   \\ \hline
loc             & \begin{tabular}[c]{@{}l@{}}It is a popular software metric \\ that determines the size of a \\ software program, by counting\\ all of the lines \\ in the program's code.\end{tabular}                                                                                                                                                  \\ \hline
lloc            & \begin{tabular}[c]{@{}l@{}}Lloc (Logical lines of code) is usually \\ used to count "statements," but their \\ meanings can change based on \\ the selection of a language.\end{tabular}                                                                                                                                              \\ \hline
sloc            & \begin{tabular}[c]{@{}l@{}}SLOC is a measure of the number \\ of lines in the source code of a \\ program, including comment lines and \\ occasionally blank lines.\end{tabular}                                                                                                                                                                           \\ \hline

comments        & \begin{tabular}[c]{@{}l@{}}This is a measure of the \\ total number of comment lines.\end{tabular}                                                                                                                                                              \\ \hline
multi           & \begin{tabular}[c]{@{}l@{}}Count of the number of \\ multi-line strings lines.\end{tabular}                                                                                                                                                                                                                                             \\ \hline
blank           & \begin{tabular}[c]{@{}l@{}}Count of the number of blank lines \\ (or whitespace).\end{tabular}                                                                                                                                                                                                                                                \\ \hline
single comments & \begin{tabular}[c]{@{}l@{}}This represents a software repository's \\ total number of unique comments.\end{tabular}                                                                                                                                                                                                                 \\ \hline
mi              & \begin{tabular}[c]{@{}l@{}}The MI (Maintainability Index) is a \\ sophisticated software indicator that \\ gauges how easily supportable \\ and modifiable the source code is. \\ The SLOC (Source Lines Of Code), \\ Cyclomatic Complexity, and Halstead \\ volume are considered in the \\ maintainability index calculation.\end{tabular} \\ \hline

Halstead Metrics\cite{halstead} & \\ \hline

h1           & distinct operators per file    \\ \hline
h2           & distinct operands per file    \\ \hline
n1           & total operators per file    \\ \hline
n1           & total operands per file    \\ \hline
Program vocabulary & vocab =n1+n2 \\ \hline
Program length  &  length =N1+N2 \\ \hline
Calculated program length & length  =N1 log(2)N1 + N2log(2)N2  \\ \hline
Volume & V=Nlog2n  \\ \hline
Difficulty & difficulty = n12⋅N2n2  \\ \hline
Effort &  effort = D⋅V  \\ \hline
Time required to program & time = E/18 seconds  \\ \hline
Number of delivered bugs &  bugs = V / 3000.  \\ \hline
\end{tabular}
\end{table}

\begin{table}[ht]
\renewcommand{\arraystretch}{1.3}
\caption{Metrics calculated from Lizard}
\label{Lizard}
\begin{tabular}{|l|l|}
\hline
\textbf{Metric} & \textbf{Description}                 \\ \hline
nloc            & Total number of lines of code without comments \\ \hline
Average NLOC    & The average number of lines per file     \\ \hline
Average CCN     &\begin{tabular}[c]{@{}l@{}} An average cyclomatic \\ complexity number\end{tabular} \\ \hline
function count  & Total parameter count of functions.        \\ \hline
Average token   & The average number of tokens per file    \\ \hline
% Google Scholar citations    &\begin{tabular}[c]{@{}l@{}} Total number of Google Scholar \\ citations for paper \\ associated with repositories \end{tabular}\\ \hline
% Semantic Scholar citations    &\begin{tabular}[c]{@{}l@{}} Total number of Semantic Scholar \\  citations for paper \\ associated with repositories \end{tabular} \\ \hline

\end{tabular}
\end{table}

\begin{figure}[htp]
    \centering
    \includegraphics[width=8.5cm]{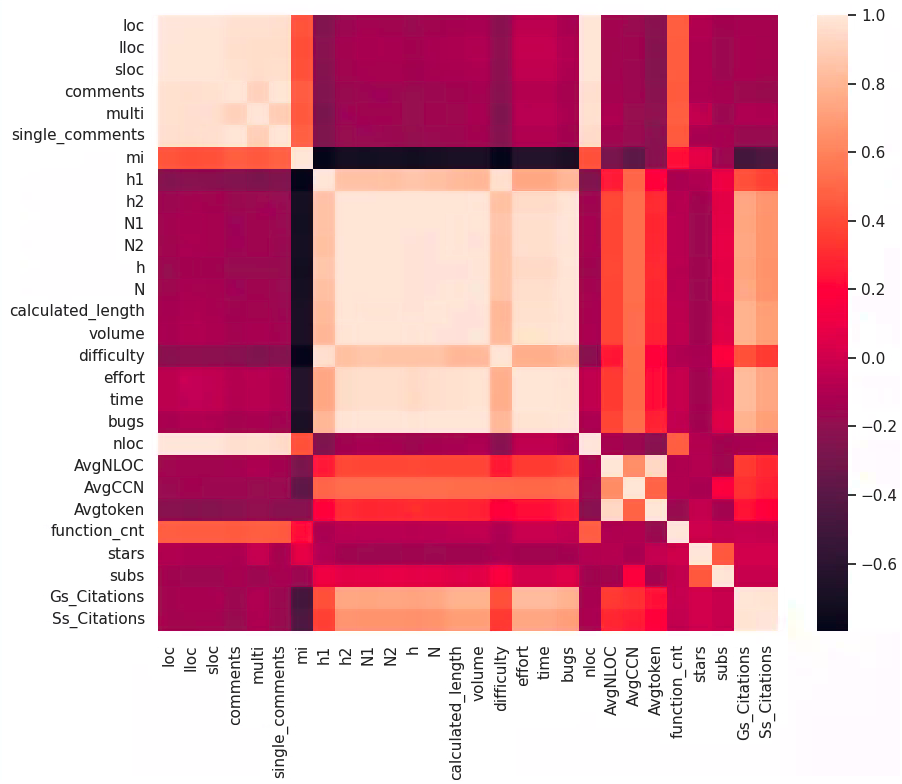}
    \caption{Correlation Heatmap of Calculated Metrics}
    \label{fig:corr}
\end{figure}

\subsection{Experiment}
Once we gathered the per-repository data in a tabular format, we conducted various exploratory data analyses to address the research questions. We then performed feature selection among the metrics using Recursive Feature Elimination (RFE) methods from the Sklearn library to generate a parsimonious model from the complete set of potential regressors that would allow us to draw conclusions about the relationship between some of the metrics and measures of popularity (i.e., stars, subscribers, and citations). With 20 possible explanatory variables, many of which were highly collinear with one another (e.g., lines of code, logical lines of code, and average lines of code), we utilized the RFE approach to examine progressively smaller subsets of variables that exhibited the highest correlations with the target variables (i.e., stars and citations). 

We also noted that the maintainability index was an approximately normally distributed variable. Ultimately, after settling on the five variables identified by the RFE method (i.e., maintainability index, lines of comments, Halstead volume, effort, and bugs), we employed Statsmodels ols methods to construct a linear model to measure the effects of the predictors. We also incorporated some robust (Huber) models after considering some of the outliers in the data, which enabled us to examine the sensitivity of the models to outliers in certain predictors. The robust linear models yielded more consistent results and reduced the impact of large outliers in the independent and dependent variables. Thus, we present the robust models in Table \ref{regression}. 

\subsection{Software Tools Used}
For our development, we used python 3.8 combined with PyGithub, radon, lizard, pandas, sweetviz, pandas\_profiling, statsmodels, sklearn, semanticscholar, scholarly, and seaborn as the main 3rd party libraries for data analysis, modeling, and visualization. 
%The full list of installed python libraries used along with versions is contained in the requirements.txt file in our GitHub repository: https://github.com/jtduran/CS573Final/requirements.txt. We also used ChatGPT to generate some LaTeX code. 

%% file: result.tex
% %Results
\begin{table*}[t]
\begin{center}
\label{data_statistics}
\caption{Summary Data Statistics of Code Metrics and Popularity Measures}

\begin{tabular}{lrrrrrrrrrrrrr}
% \begin{tabular}{l|r|r|r|r|r|r|r|r|r|r|r|r|r|}
%\toprule
\hline
{} &     loc &    lloc &    sloc &  comments &  multi &  blank &   mi &   h1 &    h2 &    N1 &    N2 &     h &     N \\
\hline
%\midrule

count &    42.0 &    42.0 &    42.0 &      42.0 &   42.0 &   42.0 & 42.0 & 42.0 &  42.0 &  42.0 &  42.0 &  42.0 &  42.0 \\
mean  &  5246.2 &  2990.8 &  3362.3 &     464.4 &  557.7 &  881.3 & 58.5 &  6.3 &  43.4 &  32.2 &  60.6 &  49.7 &  92.8 \\
std   &  7952.6 &  3968.9 &  4657.2 &     735.3 & 1392.8 & 1273.2 & 12.3 &  2.5 &  40.9 &  30.5 &  58.1 &  43.0 &  88.6 \\
min   &   222.0 &   201.0 &   194.0 &       5.0 &    0.0 &   23.0 & 19.5 &  2.8 &   8.1 &   6.9 &  13.0 &  10.9 &  19.9 \\
25\%   &  1348.2 &   920.0 &   994.2 &      81.0 &   21.2 &  209.0 & 50.8 &  4.7 &  25.5 &  18.2 &  34.7 &  30.1 &  53.0 \\
50\%   &  2689.0 &  1701.0 &  1913.0 &     165.0 &   39.0 &  489.5 & 58.4 &  5.8 &  29.6 &  21.5 &  39.0 &  34.8 &  60.5 \\
75\%   &  3852.5 &  2564.8 &  2625.8 &     397.5 &  240.5 &  764.0 & 67.0 &  7.0 &  44.9 &  34.5 &  65.0 &  51.5 & 100.7 \\
max   & 34769.0 & 17595.0 & 20763.0 &    3025.0 & 5682.0 & 5532.0 & 79.0 & 14.2 & 256.5 & 191.2 & 364.2 & 270.8 & 555.5 \\
\hline
\hspace{3em}
%\bottomrule
\end{tabular}

\begin{tabular}{lrrrrrrrrrrr}
%\toprule
\hline
{} &  calculated\_length &  volume &  difficulty &  effort &   time &  bugs &    nloc &  AvgNLOC &  AvgCCN &  Avgtoken &  function\_count \\
%\midrule
\hline
count &               42.0 &    42.0 &        42.0 &    42.0 &   42.0 &  42.0 &    42.0 &     42.0 &    42.0 &      42.0 &          42.0 \\
mean  &              296.7 &   612.4 &         4.4 &  5376.6 &  298.7 &   0.2 &  3705.4 &     15.9 &     3.1 &     148.8 &         110.9 \\
std   &              353.7 &   752.9 &         1.8 &  9520.7 &  528.9 &   0.3 &  5778.3 &      8.7 &     1.3 &      93.2 &         214.5 \\
min   &               40.1 &    90.1 &         2.1 &   520.8 &   28.9 &   0.0 &   194.0 &      7.0 &     1.7 &      69.5 &           4.0 \\
25\%   &              151.1 &   288.1 &         3.2 &  1784.4 &   99.1 &   0.1 &   931.8 &     10.2 &     2.2 &      90.8 &          10.9 \\
50\%   &              182.0 &   389.8 &         4.0 &  2809.0 &  156.1 &   0.1 &  1823.0 &     13.8 &     2.9 &     131.2 &          52.0 \\
75\%   &              292.8 &   640.2 &         5.0 &  4825.1 &  268.1 &   0.2 &  2752.8 &     17.8 &     3.3 &     162.2 &         112.5 \\
max   &             2232.8 &  4761.7 &        10.1 & 60420.1 & 3356.7 &   1.6 & 25490.0 &     49.5 &     7.2 &     579.5 &        1327.0 \\
\hline
\hspace{3em}
%\bottomrule
\end{tabular}

\begin{tabular}{lrrrr}
\hline
{} &  stars &  subscribers &  google scholar &  semantic scholar \\
\hline
%\midrule
count &   42.0 &  42.0 &          42.0 &          42.0 \\
mean  &   24.6 &   2.5 &          57.8 &          47.3 \\
std   &   41.4 &   1.3 &         155.9 &         121.1 \\
min   &    2.0 &   1.0 &           0.0 &           0.0 \\
25\%   &    5.0 &   2.0 &           6.0 &           5.0 \\
50\%   &   16.9 &   2.0 &          13.0 &          13.5 \\
75\%   &   30.0 &   3.0 &          44.2 &          34.5 \\
max   &  264.0 &   7.0 &         976.0 &         702.0 \\
\hline
%\bottomrule
\end{tabular}
\end{center}
\end{table*}

\begin{table}
\caption{ Robust Regression Results of Popularity measure with Code Metrics. }
\label{regression}
\begin{center}
\resizebox{\columnwidth}{!}{%
\begin{tabular}{lllll}
\hline
 & subscribers     & stars    & google scholar & semantic scholar  \\
\hline
mi & 0.033*** & 0.392*** & 0.538**       & 0.438**        \\
         & (0.006)  & (0.092)  & (0.226)       & (0.181)        \\
comments & -0.000   & -0.005   & -0.029***     & -0.022***      \\
         & (0.000)  & (0.004)  & (0.010)       & (0.008)        \\
volume   & 0.002    & -0.009   & -0.241***     & -0.175***      \\
         & (0.001)  & (0.020)  & (0.049)       & (0.039)        \\
effort   & -0.000   & 0.001    & 0.034***      & 0.025***       \\
         & (0.000)  & (0.002)  & (0.004)       & (0.003)        \\
bugs     & 0.000    & -0.000   & -0.000***     & -0.000***      \\
         & (0.000)  & (0.000)  & (0.000)       & (0.000)        \\
\hline
*** significant at 1\% \\
**  signification at 5\% \\
\end{tabular}
}
\end{center}
\end{table}

Here we examine the results of our analysis of the various code metrics, present the results of the regressions between the metrics, citations, and GitHub measures, and our ideas about the underlying reasons for the observed relationships.

\textbf{RQ1: Is there any correlation between different code metrics and research citations?}

According to Table~\ref{regression}, all the predictors, including mi (maintainability index), comments (number of lines of comments), volume (Halstead volume), effort (Halstead effort), and bugs (Halstead bugs), exhibit statistically significant association with both Google Scholar and Semantic Scholar citations in the robust model. Specifically, lines of comments, bugs, and code volume demonstrate a negative correlation with both gs and ss citations, while mi and Halstead effort exhibit a positive correlation. Interestingly, these findings align with our intuition, indicating that code that is more concise, less buggy, and less commented is linked to fewer paper citations. In contrast, more maintainable, high-effort code is associated with more citations.

\textbf{RQ2: Is there any correlation between different code metrics and GitHub popularity?} 

As mentioned previously, the measures of GitHub stars and subscribers are considerably more erratic and possibly prejudiced when compared to citations. Consequently, we find the associations between these measures and the various code metrics to be less compelling. According to Table~\ref{regression}, the only statistically significant relationship is with mi (the Maintainability Index), which displays a positive association with both stars and subscribers. While comments are not statistically significant, they exhibit a negative association with the GitHub measures, which is widespread across all the measures. The remaining Halstead metrics, however, appear to have coefficients of varying signs between stars and subscribers and citations, indicating no persistent association.

\textbf{RQ3: Is there any correlation between GitHub popularity and associated paper citations regarding code metrics?} 

Table~\ref{regression} and the correlation heat map presented in Fig.~\ref{fig:corr} both demonstrate a strong, statistically significant, positive correlation between the Maintainability Index and stars, subscribers, as well as both Google Scholar and Semantic Scholar citations. This was the most robust and consistent finding throughout our study, exhibiting significance at the 5\% and higher level across all models. Furthermore, all four models consistently showed the signs of coefficients and occasional significance of regressors. Interestingly, the most unusual discovery was the persistent negative correlation between the number of comments and code popularity (across GitHub and citations), which could indicate that high-quality code documents itself, as per the old adage. 

Overall, we are satisfied with the general level of intuitiveness of the results and the simplicity of interpretability of the coefficients. While it does not suggest any underlying causality, we believe it is a strong indication of a few intuitive measures that reflect high-quality code and research, which are more likely to be rebuilt upon and reused to advance the science of fake news detection. 

%% file: threats.tex
Our focus was solely on repositories that contained published papers. A simple search for fake news on the GitHub platform yielded over 12,000 repositories. However, we narrowed our search to repositories that aligned with a publication in order to identify the best available tools or codes in the fake news domain. Most repositories associated with fake news were either class projects or work without any related publication. As a result, our conclusion may not be the most comprehensive or generalizable from the limited repositories we found. Nevertheless, we have included the most popular repositories and tools associated with fake news, which provide a good understanding of the topic. 

We sourced most of the repositories for our analysis from the GitHub platform. Still, other popular code hosting platforms, such as Bitbucket and GitLab, may contain more robust tools in this domain. This study did not consider proprietary industry-level software or tools, particularly those developed by large companies and used in conjunction with search engines. However, since GitHub is the most widely used public code hosting platform, it largely satisfied the need to evaluate similar repositories on other code hosting platforms. We found this threat to be minimal. Furthermore, our study focused solely on publicly available and easily accessible free tools. We did not consider paid or large-scale industry-based tools within the scope of this research.    

Our study incorporated several research techniques and features. However, conducting a more comprehensive analysis, such as surveying users, could provide further insight into their experience with fake news detection tools. To assess the effectiveness of existing tools, we utilized popularity measures such as stars, followers count, and forks. An expanded analysis with additional features could potentially provide more information on whether popularity is a reliable indicator of the best code, particularly in the context of fake news detection.

%% file: conclusion.tex
%Conclusion
In recent years, fake news has emerged as a major concern. One approach to combat the spread of fake news is to create and distribute high-quality research associated with identifying and analyzing misinformation. A crucial aspect of this effort is sharing research code repositories. By identifying important characteristics of this code, we can gain valuable insights into improving the research quality.

This paper presents code metrics and code repository analysis of fake news studies to better understand the characteristics of the most popular and highly cited code and research. We also advocate source code quality and associated measures of repository popularity. Through this analysis, we uncover several significant findings.

Our analysis shows that a typical research project repository contains approximately 2000 lines of code, a few hundred lines of comments, and a maintainability index of approximately 50, which falls in the middle of the 1 to 100 scale. In addition, a typical source code file is relatively straightforward, with only a few independent code paths, likely contributing to its increased maintainability.

We found that papers associated with repositories exhibit a strongly right-skewed distribution regarding citations, with seminal papers in the area or those published in well-known venues like Nature being more heavily cited. In contrast, measures of popularity such as GitHub stars or subscribers exhibit weaker associations with the code metrics we measure than citations. We also observed that citations from both Google Scholar and Semantic Scholar are highly similar, with Google Scholar having an average of 22\% more citations. Furthermore, the code metric measures and popularity measures represent a mixture of normally distributed measures and those with many outliers on both the high and low ends of the distributions, which complicates modeling and needs further attention in cleaning the data set. 

Overall, we found that more popular fake news repositories and associated papers with higher citation counts tend to have more maintainable code measures, more complex code paths, a larger number of lines of code, a higher Halstead effort, and fewer comments. 

In summary, our analysis of code repositories provides valuable insights into the characteristics and quality of fake news repositories on GitHub. These findings can hopefully inform efforts to identify and track fake news, facilitate the development of research in this area, and ultimately help mitigate the negative impact of this information on individuals and society.

%In conclusion, fake news is a major concern that can significantly negatively impact individuals and society. Source Code repository analysis provides valuable insights into the characteristics and behavior of fake news research projects that help to mitigate this problem and can inform efforts to identify and track misinformation. By using these findings to develop effective research and coding strategies to combat fake news, we can work towards a more informed and better-informed society.